# Three dimensional simulations of embolic stroke: an equation for sizing emboli from imaging

James P. Hague, PhD
School of Physical Sciences, The Open University, Walton Hall, Milton Keynes UK, MK7 6AA
Corresponding author: Jim.Hague@open.ac.uk, +44 1908 654626

Jonathan Keelan, PhD
School of Physical Sciences, The Open University, Walton Hall, Milton Keynes UK, MK7 6AA

Lucy Beishon, MD, PhD
Department of Cardiovascular Sciences, University of Leicester, Leicester UK, LE1 7RH

David Swienton, MD
Department of Radiology, University Hospitals of Leicester NHS Trust, Leicester UK, LE1 5WW

Thompson G Robinson, MD
University of Leicester, Department of Cardiovascular Sciences, Leicester UK, LE1 7RH
NIHR Leicester Biomedical Research Centre, British Heart Foundation Cardiovascular Research Centre, Leicester UK, LE5 4PW

Emma M.L. Chung, PhD
Department of Cardiovascular Sciences, University of Leicester, Leicester, UK, LE1 7RH
University Hospitals of Leicester NHS Trust, Department of Medical Physics, Leicester Royal Infirmary, Leicester, UK, LE1 5WW
School of Life Course and Population Sciences, King's College London, Guy's Campus, SE1 1UL

# Abstract

There is a need to develop Monte Carlo simulations of stroke to run in-silico trials to replace animal models, explore clinical scenarios to develop hypotheses for clinical studies and for interpreting clinical monitoring. We perform three-dimensional (3D) stroke simulations, carrying out in-silico trials to relate lesion volume to embolus diameter and calculate probabilistic lesion overlap maps, building on our previous Monte Carlo method. Simulated emboli were released into a 3D *in silico* vasculature, supplying gray and white matter brain volumes, to generate individual lesion estimates and probabilistic lesion overlap maps. Computer generated lesions were assessed by clinicians and compared with real world radiological images. Simulations of large single emboli reproduced similar middle cerebral artery (MCA), posterior cerebral artery (PCA) and anterior cerebral artery (ACA) lesions to those observed clinically. A proof-of-concept in-silico trial led to a conjecture relating estimated infarct volume as a percentage of total brain volume to relative embolus diameter: $relative\ diameter\ =\ [\%\ infarct\ volume\ /\ a]^{1/b}$ where a= 104.2 +/- 0.98, b=3.380 +/- 0.030. Probabilistic lesion overlap maps were created, confirming the MCA territory as the most probable resting place of emboli in the computational vasculature, followed by the PCA then ACA. The article shows proof of concept for developing a 3D stroke model from an automatically constructed vasculature.

# 1. Introduction

The consequences of emboli entering the cerebral vasculature can be devastating, and given the wide range of origins for embolic stroke, a universal numerical model that links disparate patterns of embolisation to stroke outcomes would be transformative to stroke research. Cardioembolic stroke (typically caused by a small number of large solid emboli) results in severe stroke presentations, and the incidence is projected to triple by 2050 [1]. Large emboli can also form from other sources (e.g. extracranial atherosclerosis) and also cause severe ischemia [2]. There are several situations where large numbers of emboli can be formed, such as during cardiac surgery [3,4,5]; carotid surgery [6]; diving decompression [7]; and from mechanical heart valves [8] which can lead to varying risks of neurocognitive decline and stroke.

There is a need to develop numerical models of embolic stroke to run in-silico trials to replace animal models, explore clinical scenarios to develop hypotheses for clinical studies, and provide interpretation for intraoperative monitoring. Numerical models of stroke have various potential benefits, including the potential to run in-silico "clinical" trials, partially replace animal models, explore medical scenarios (e.g. embolus fragmentation, decompression sickness and embolisation during surgical procedures), and provide complementary information about the role of embolisation in parts of the cerebral vasculature that are inaccessible to imaging. This is important because results from animal studies of stroke may not be broadly applicable to human trials [9], recruitment to clinical trials may be challenging, and imaging techniques have limited resolution. By testing medical scenarios, computational (in-silico) models can be used to provide and assess initial hypotheses for clinical studies, potentially saving time and resources allocated to clinical trials that have no prospect of success. In this paper we demonstrate how 3D stroke simulations can be used to determine lesion volumes and locations from embolus properties, and perform an in-silico study leading to a conjectured relationship between lesion volume and embolus diameter.

Various numerical methods have previously been used to understand the motion of emboli through the cerebral vasculature. On the large vessel scale, the motion of emboli through various circle of Willis (CoW) variations has been studied using computational fluid dynamics (CFD), neglecting small vessels or embolus-flow interactions [10,11]. On the cellular scale, models including ion channels, cell metabolism and apoptosis have been used to study penumbra and lesion evolution following stroke [12,13,14]. Porous circulation models representing the smallest vessels have been combined with one-dimensional simulation of flow in the largest arteries of the brain (obtained from imaging, with a minimum vessel diameter of approximately 0.3 mm), within which a stroke is simulated by pinching off vessels *in silico* [15,16] (in such models arterioles are treated very differently to arteries). We previously introduced a Monte Carlo method with embolus-flow interactions, but without spatial information, for predicting the severity of strokes from specific embolisation patterns [17,18,19,3]. Limited spatial information has been included in a similar stroke model using a tree generated with constrained constructive optimisation (details of the vasculature and its shape were not provided) [20].

The goal of this paper is development of a 3D Monte Carlo simulation of stroke that includes an automatically generated brain vasculature which contains vessels of all scales (from the mm scales of the largest arteries to the micron scales of the smallest arterioles) and which can mimic large cohort studies. This 3D *in silico* stroke model tracks emboli as they move through the vasculature to determine the locations at which they generate lesions. The Monte Carlo nature of the simulation makes it possible to run in-silico studies that mimic large cohort or population studies. As examples we run in-silico studies leading us to conjecture a simple formula for determining the diameter of the embolus causing a lesion of specific volume, and to construct probabilistic lesion overlap maps. This could form a useful basis for future '*in silico*' clinical trials aimed at investigating the impact of clinical interventions on the incidence, volume and location of new lesions or quantifying epidemiological population-based trends in stroke imaging and outcomes over time.

## 2. Methods

### 2.1 Monte Carlo simulations

A Monte Carlo method was used to simulate strokes. Our method was previously used with a symmetric vasculature that lacked spatial information [17,18,19,3]. We have extended our Monte Carlo method with an *in silico* vasculature [21, 22]. The *in silico* vasculature is grown using the simulated annealing vascular optimisation (SALVO) technique [23,21,24] and consists of 8192 segments representing vessels connected by bifurcations, grown according to the distribution of gray and white matter from images of a healthy individual, and with input vessel situated at the approximate location of the circle of Willis (of diameter $d_0$ related to diameters of middle cerebral artery, MCA, posterior cerebral artery, PCA, anterior cerebral artery, ACA and cerebellar artery (CA) via $d_0^\gamma = d_{MCA}^\gamma + d_{PCA}^\gamma + d_{ACA}^\gamma + d_{CA}^\gamma$, where $\gamma = 3.2$) [21]. In this paper, all vessels in the in-silico vasculature have been scaled such that the input vessel of the in-silico generated arterial tree is consistent with typical diameters of the major cerebral arteries (taken as $d_{MCA} = 3.1mm$ for MCA [25], $d_{PCA} = 2.7mm$ for PCA [26], $d_{ACA} = 2.6mm$ for ACA [27] and $d_{CA} = 1.5mm$ for CA [28]), leading to $d_0 = 4mm$. We have symmetrized the vasculature about the sagittal plane to create a 16384 segment tree. The generated vasculature contains three major arteries, approximating the MCA, PCA and ACA perfusion territories. Figure 1 shows the computationally generated vasculature alongside its resulting perfusion territories [21]. We note some differences between the perfusion territories of the *in silico* vasculature and the human brain. In the *in silico* model, PCA and cerebellar territories are fused and the MCA territory is larger than typical. In terms of in-silico trials, this is not expected to cause problems, however a closer reproduction would be needed to perform inverse problems. We note a large branch from the in-silico PCA that supplies a region of the cerebellum and may act as a proxy for the inferior cerebellar arteries. To each leaf (terminal) vessel of this symmetrized vasculature we add a symmetric vascular tree with 32 elements and bifurcation exponent $\gamma = 3.2$, such that the minimum vessel diameters are 14.9 microns. The root vessels of the two hemispheres are joined at a single input for convenience. Such a vasculature has vessel sizes on all scales down to the capillary scale, and the ability to treat the metabolic requirements of gray and white matter separately. Such a vascular tree is consistent with a complete circle of Willis, which applies to approximately 50% of the population (CoW completeness can be determined by scan).

Into this tree we place simulated emboli which move until they either exit the vasculature or block a vessel. At each bifurcation, an embolus passes down one of the two branches with a probability decided by the relative flow in each direction according to a Monte Carlo scheme, unless (a) one of the branches has a smaller diameter than the embolus, in which case it travels down the larger branch (which we consider more likely due to forces on the embolus from the flow) or (b) both branches have a smaller diameter than the embolus, in which case the parent vessel is obstructed. The ability for emboli to travel different paths reflects the complexity of fluid dynamics, which is highly sensitive to initial conditions. Leaf vessels of the whole tree are monitored for lack of flow (ischaemia), and a lesion is deemed to have formed if no flow has been received for a simulated period of 4 hours. We explore two embolus dissolution rates: a reduction of 0.533mm diameter per 24 hours (5 days for a 2.67mm diameter embolus) and a reduction of 0.267mm diameter per 24 hours (10 days for a 2.67mm diameter embolus). Results from the model are insensitive to such large differences in dissolution rate especially for large emboli, the key factor being that the dissolution rate for a large embolus is much greater than the timescale for ischaemia. By running the model with various embolus sizes and Monte Carlo sequences, we run in-silico studies mimicking large cohort studies.

## 2.2 Simulation classification

LB reviewed 50 simulation images, provided in random order of size, and classified them according to whether they would be likely to be observed in a clinical setting (follow up clinic) and assigned a characterisation of 'yes', 'no' or 'rare'. LB was blinded to embolus size. Results were then binned into histograms according to embolus size.

## 2.3 Image selection

Meetings were undertaken by specialists in neuroradiology (DS), stroke and geriatric medicine (TGR, LB). The Anatomical Tracings of Lesions After Stroke (ATLAS) dataset (a large open access database of 304 manually segmented T1-weighted MRI stroke images [29]) was reviewed by one researcher (LB). Typical examples of anterior, middle and posterior territory embolic stroke were selected from the database. Manual segmentation is currently the gold standard method for segmenting brain lesions on T1-weighted imaging. Images were selected for this study by consensus discussion between the three Clinicians to ensure they were representative of embolic lesions and to confirm the perfusion territory affected. Cases were chosen from the ATLAS data-set that were representative of the following radiological characteristics associated with embolic stroke: multiple vascular territories and or hemispheres affected, cortical, cerebellar, or brainstem location, >1.5 cm diameter lesions [30,31][1]. These images were reviewed at a consensus meeting held between LB, TGR and DS and the best examples for each territory were selected for replication by the stroke simulation based on the same criteria. Initial simulated images were compared at a further consensus meeting against the ATLAS dataset. Qualitative feedback was obtained with regard to the size of the simulated perfusion territories and limitations of the model and the number of sample simulations was increased to 1000 before matching to the clinical images, resulting in a final set of simulations for review by the consensus group. Clinical feedback from the original simulated images was used to identify matching images from a larger number of simulations.

---

[1] Multiple infarcts will be discussed in a future publication.

## 3. Results

The majority of stroke outcomes from the model for realistic embolus sizes appear clinical. Figure 2 shows assessment of simulated strokes by a Clinician as to whether they resembled lesions seen in a follow up clinic, finding that lesions in the in-silico model resembling those in clinic or rarely seen in clinic are typically caused by emboli with diameters of 2.85 mm or less. 100% of simulated lesions caused by emboli with diameters <2.25mm were rated as "typically seen in clinic". Simulated lesions identified as similar to those typically seen in clinics related to a maximum embolus diameter of 2.85 mm. Some *in silico* lesions caused by large emboli of between 2.25 and 3.15 mm diameter were classified as rarely seen in clinic, and accounted for ~4% of simulated lesions. Most *in silico* lesions generated by emboli of 2.85 mm or greater were not consistent with those typically seen in clinic (as expected).

Simulations featuring emboli of varying size entering the *in silico* vasculature show that infarct volume grows rapidly with embolus diameter, following a power-law distribution as shown in Figure 3. The infarct volume as a percentage of total brain volume fits well to a power law given by:

$$\% \; infarct \; volume \; = \; a(d/d_0)^b \qquad \text{(Equation 1)}$$

where the diameter of the embolus is *d*. The dimensionless parameters $a = 104.2 \pm 0.98$, and $b = 3.380 \pm 0.030$, The reason for providing this result as a percentage of total brain volume and $d/d_0$ is that the expression is independent of variations in individual brain volume and vessel diameters. Equation 1 can be inverted to determine the diameter of an embolus causing an infarct as:

$$d = d_0 \times [\% \; infarct \; volume \; / \; a]^{1/b} \qquad \text{(Equation 2)}$$

(we note that this equation provides a conjecture / hypothesis and that clinical studies would be needed to confirm this relationship). Variations in the infarct volume occur because emboli of similar sizes may travel different paths due to the Monte Carlo nature of the simulations, and were typically less than a factor of two. Lesion volumes were calculated for two different dissolution rates, and results were insensitive to this parameter.

Specific instances of stroke are presented in Figure 4. The left hand panels of Figure 4 show examples of stroke from the ATLAS database, and the right hand panels show examples of similar strokes generated by our computational model. In each computational case, a single embolus was introduced into the vascular tree and allowed to propagate through the arterial tree. The results have been compared with scans from the ATLAS database that were confirmed to be representative of embolic stroke by the specialist clinicians (TGR, DS, LB). Strokes occurred in the MCA, PCA and ACA territories of the *in silico* vasculature. We note that ACA strokes were particularly rare. Qualitative feedback from the clinical consensus meeting was that: (1) Smaller MCA strokes were well replicated in the simulation images. The largest MCA strokes were too extensive, involving other perfusion territories (ACA and PCA), consistent with results in Figure 2. (2) ACA stroke in some of the simulated images included an area affecting the MCA region which was not typically seen clinically and may be due to the mis-match between *in silico* and human perfusion territories. (3) PCA stroke was well replicated in the simulated images. (4) Borderzone infarcts were less well replicated in terms of location and shape of the infarct. These differences in territory likely match those of the in-silico vasculature, and we do not expect them to lead to major problems if used for hypothesis generation. Taken in the context of Figure 2, there is good reproduction up to a certain size, beyond which emboli become so large that they block multiple territories leading to a mis-match.

Probabilistic lesion overlap maps showing spatial distributions of lesions resulting from different size distributions of emboli are shown in Figure 5. These maps provide an estimate of the risk that an embolus causes a lesion at a specific brain location. The spatial distribution of lesions was found to be highly dependent on embolus size. Figure 5 shows the probability of lesions for (a) emboli of 0.8 mm diameter, (b) emboli of 1.6 mm diameter, and (c) emboli of 2.4 mm diameter. Of these three sizes, the 0.8mm diameter emboli lead to the most spatially homogeneous distribution of lesions across the brain affecting multiple perfusion territories. 2.4mm diameter emboli preferentially block the MCA territory. 1.6mm diameter emboli preferentially reach the MCA and PCA territory, with some ACA territory strokes. Thus, the size of emboli influences their trajectories through the vasculature and indicates a preference for larger emboli traveling into the MCA territory, leading to larger lesions occurring in the MCA territory.

When a flat distribution featuring equal numbers of emboli of differing sizes, up to a 2.67 mm maximum diameter is introduced to the *in silico* vasculature, obstruction of the posterior MCA territory was most common, with further blockages occurring in both the PCA territory, and the remainder of the MCA territory, but with fewer lesions in the ACA territory (Figure 5(d)). A flat distribution is used because the true distribution of embolus sizes relating to strokes seen in clinics is not known. In the ATLAS data-set, MCA infarcts are most probable. This is also true in our *in silico* vasculature, although the most probable location in the simulations was found to be towards the superior-posterior region of the MCA territory. Our simulations have large numbers of lesions in the PCA territory. ACA infarcts were rare in both the ATLAS dataset and our simulations.

## 4. Discussion

In this study, we demonstrated that 3D Monte Carlo simulations of embolic stroke can be used to make predictions about the size distributions and locations of lesions due to specific patterns of embolisation. There are several advantages to stroke modeling in this way. Firstly, spatial information is available enabling direct comparison with radiological images. Secondly, *in silico* clinical trials could be conducted. Thirdly, the simulations require no prior knowledge of the patient's anatomy or status other than that already contained within standard brain imaging data. Clinical feedback indicates that simulated lesions were realistic for embolus sizes below 2.85mm, comparable with the MCA diameter.

By way of example, we have determined an expression to relate embolus size to lesion volume. Thus, we have shown how large in-silico trials can be used to develop a hypothesis. Clinical studies need to be carried out to identify if the information obtained from this expression is useful and accurate for clinical outcomes. We have also shown how probabilistic lesion overlap maps can be generated using the numerical model. Such maps could be useful as input to Bayesian techniques.

As the sophistication of *in silico* models improves, there is an opportunity to complement or replace animal models and to extend our model based on cellular biology to perform detailed *in silico* clinical trials. Experimental models of stroke are often animal based, but may have limited applicability to humans [9]. So complementary numerical models could improve decisions on which hypotheses to take forward into clinical studies.

Another application of the stroke simulations presented here relates to multiple embolisation events which were not considered in this paper. Monte Carlo methods are ideal for simulating showers of solid and gaseous emboli during cardiovascular interventions, such as cardiac or carotid surgery, cardiac ablation and stent deployment [17,18,19,3]. In these settings, embolic events can be detected using intraoperative monitoring and could be directly incorporated into a 3D stroke simulation to predict the impact of emboli on brain tissue perfusion, in a similar manner to Ref. [3]. We will discuss the modeling of multiple embolisation in a future publication.

A strength of our simulation method is inclusion of vessels on all length scales, embolus-flow interactions and tracking of embolus paths through the tree. Further strengths include comparison with a large real-world image database which used manual lesion segmentation, with all scans and simulation images reviewed by a consultant neuroradiologist before inclusion in this study (since images in ATLAS were not always segmented by neuroradiology experts).

Future extensions to the model include: Enhancements to the vasculature, using e.g. vasculature taken from MRI imaging augmented using the SALVO method, simulations differentiating between gaseous and solid emboli, better representation of the CoW and personalisation. Refinements to CoW representation would improve stroke simulations for cases where embolus sizes exceed the diameters of the largest vessels exiting the CoW, so that lesions do not cross ACA, PCA and MCA boundaries. Personalization would enhance the in-silico vasculature by seeding the vascular growth algorithm based on patient-specific time-of-flight MR angiography data, with additional arterial inputs for ophthalmic arteries, cerebellar arteries and other major vessels, and representative variants of the CoW (to account for the 50% of the population that do not have a complete CoW, including any related modifications to perfusion territories due to CoW topology). Ideally databases containing acute rather than chronic stroke images would be used for comparison with simulation results.

Further downstream, we propose that Monte Carlo stroke simulations could be used to determine the cause of an acute ischaemic stroke. Personalized stroke simulations aimed at rapid diagnoses of the location and source of a stroke could be useful to distinguish cardioembolic sources from other types of stroke, facilitating rapid access to targeted treatment. This is a complex problem currently based on analysis of a range of factors including patient characteristics and risk factors, as well as neurological symptoms and imaging data (as the location, size, and pattern of infarcts often indicate likely etiology). At present there are no quantitative criteria or tools by which this is assessed, relying on the skills of an experienced Radiologist, or treating Clinician. [32] Up to 25% of strokes are currently classified as embolic of undetermined source (ESUS) [32, 33] and correct identification of atrial fibrillation (AF) is thought to occur in only 50% of eligible patients, representing a significant barrier to timely anticoagulation [32]. Rapid identification of the source of emboli, and ability to distinguish between cardioembolic and other causes of stroke, has the potential to guide clinical treatment and improve patient outcomes. Further tailoring of simulations to include patient-specific imaging data and match *in silico* predictions to real-world imaging, coupled with implementation of artificial intelligence (AI)-based feature extraction (segmentation) from images may be useful for automatically predicting the size, location (and potentially the source) of embolisation to improve image interpretation methods, and save neuroradiologists' time. As a first step in this direction, we propose that Equation (2) can be used as a tool to provide estimates of embolus size from imaging data, especially in cases where it is difficult to determine the lumen diameter for an occluded vessel (we note that further clinical study would be needed to validate the equation, and assess if such additional information improves patient outcomes). A limitation of our simulations for this application is that *in silico* perfusion territories reproduced by the model differ slightly from those seen in the human brain. As the simulations and their associated probabilistic lesion overlap maps (and other measures such as the correlation between the hemispheres where multiple lesions occur, or whether the lesions are subcortical) are improved, we suggest that the probability that the embolus has a particular size and type (and ultimately origin information) could be extracted by e.g. comparing imaged lesions to the probabilistic lesion overlap maps associated with different embolic sources. Additional steps in this direction should include comparison of simulation results with patient diagnosis and outcome to inform predictive diagnostics. Ultimately, machine learning methods could be used to train the model and improve its predictive capacity, and would allow the patient's symptoms, physiology, cellular biology, and genetic or demographic data, to be incorporated with advanced image interpretation.

# Funding

JK acknowledges EPSRC grant no. EP/P505046/1. LB is a clinical research fellow supported by the Dunhill Medical Trust (RTF1806\27). TGR is a National Institute for Health Senior Investigator.

Figure 1: Vessels and perfusion territories in the in silico model. Green and turquoise - ACA territories. Blue and yellow - MCA territories. Purple and red - PCA territory and cerebellum.

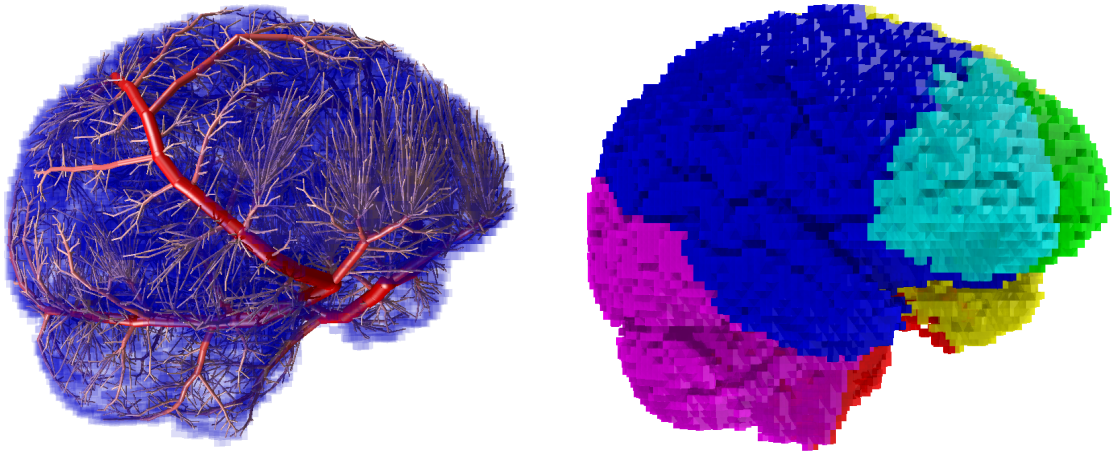

Figure 2: Clinical classification of simulated strokes. A clinician (LB) was given images of the *in-silico* lesions and asked to assess them as to whether they resemble cases relating to follow-up clinic "likely to be seen in clinic", "rare in clinic" or "not seen in clinic". This may suggest the maximum embolus diameter commonly seen in follow-up clinic is ~2.85mm.

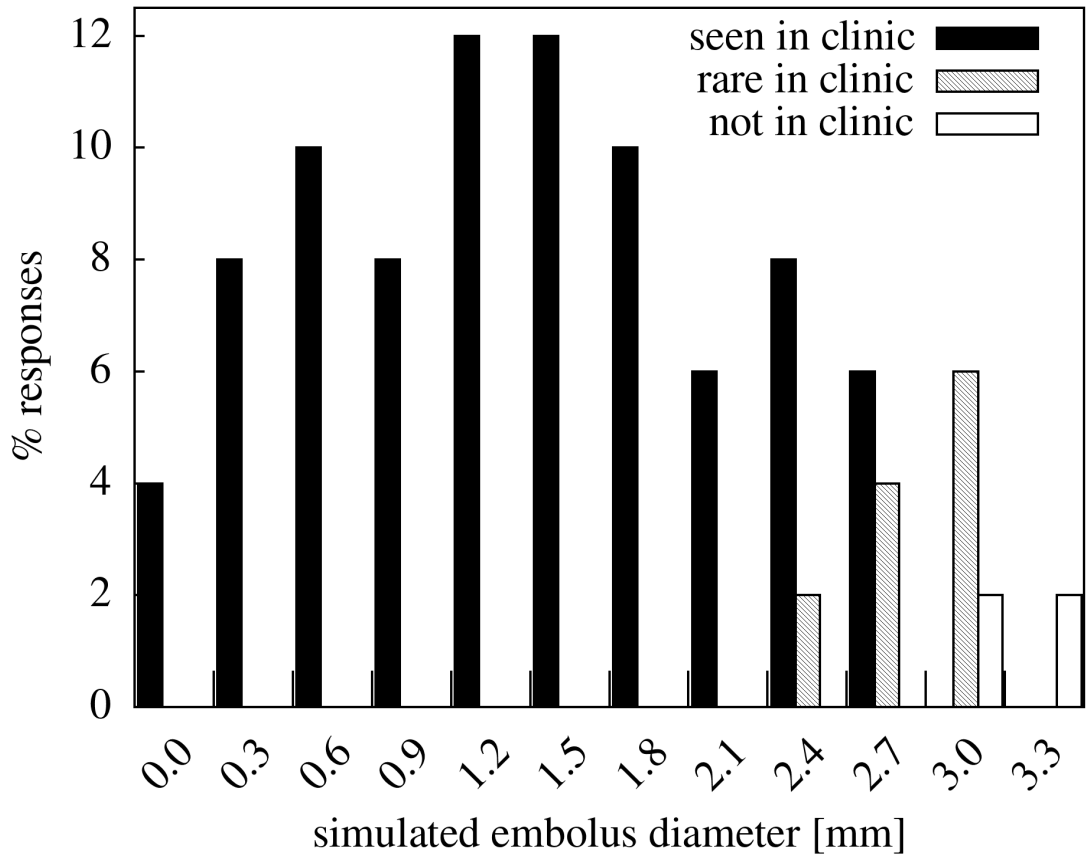

Figure 3: Volume of *in silico* infarcts as a proportion of brain size increases with embolus diameter according to a power law. Both the infarct volume and embolus diameter are plotted on logarithmic scales to clearly demonstrate this power law relationship. Lesion volumes were calculated for two different dissolution rates, and results were insensitive to this parameter.

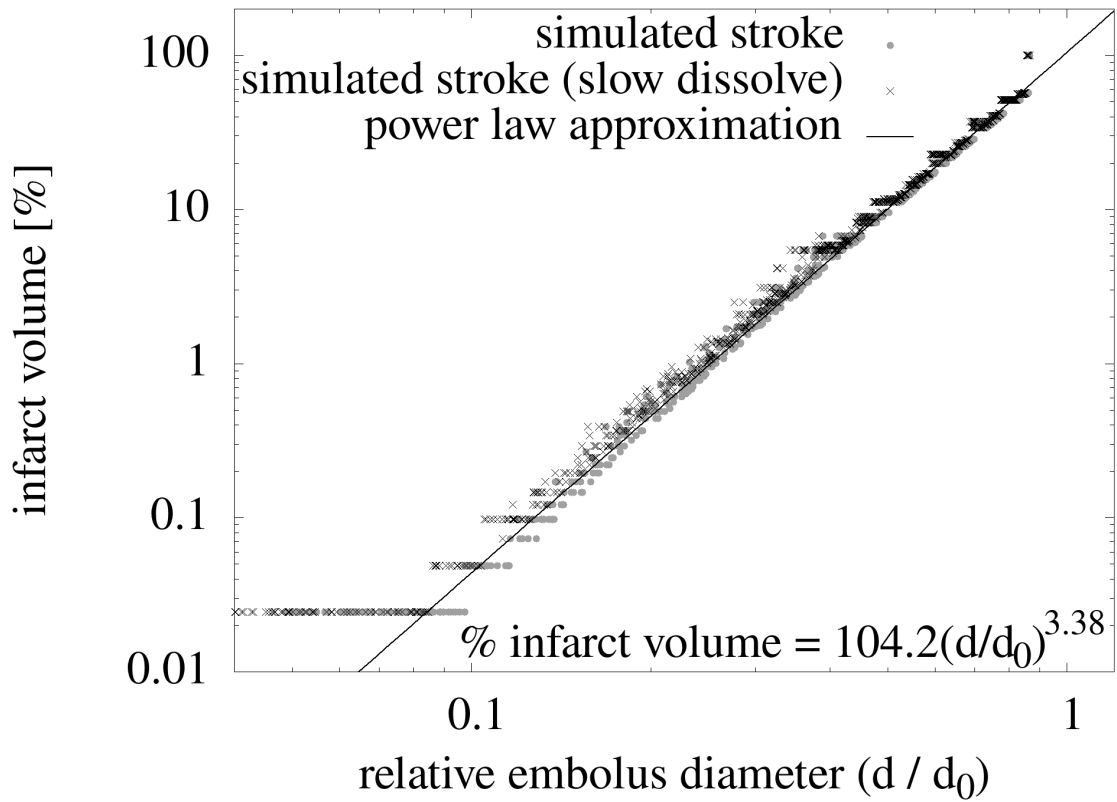

Figure 4: Stroke examples extracted from the ATLAS database, compared to an example stroke simulation for the same perfusion territory. (a) Smaller MCA strokes were well replicated in the simulation images, (b) A mismatch was observed in the position of simulated lesions in the ACA perfusion territory, which extends more superiorly from anterior to posterior in the human than in our simulations. (c) PCA territory lesions were well reproduced by the simulation.

(a) MCA stroke

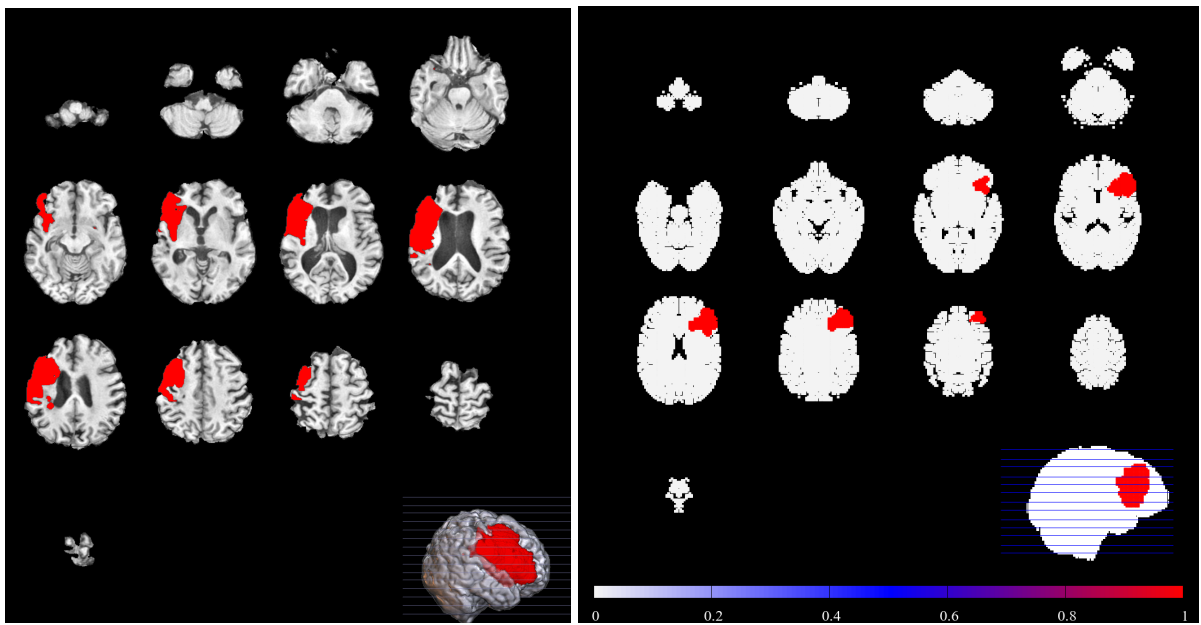

(b) ACA stroke:

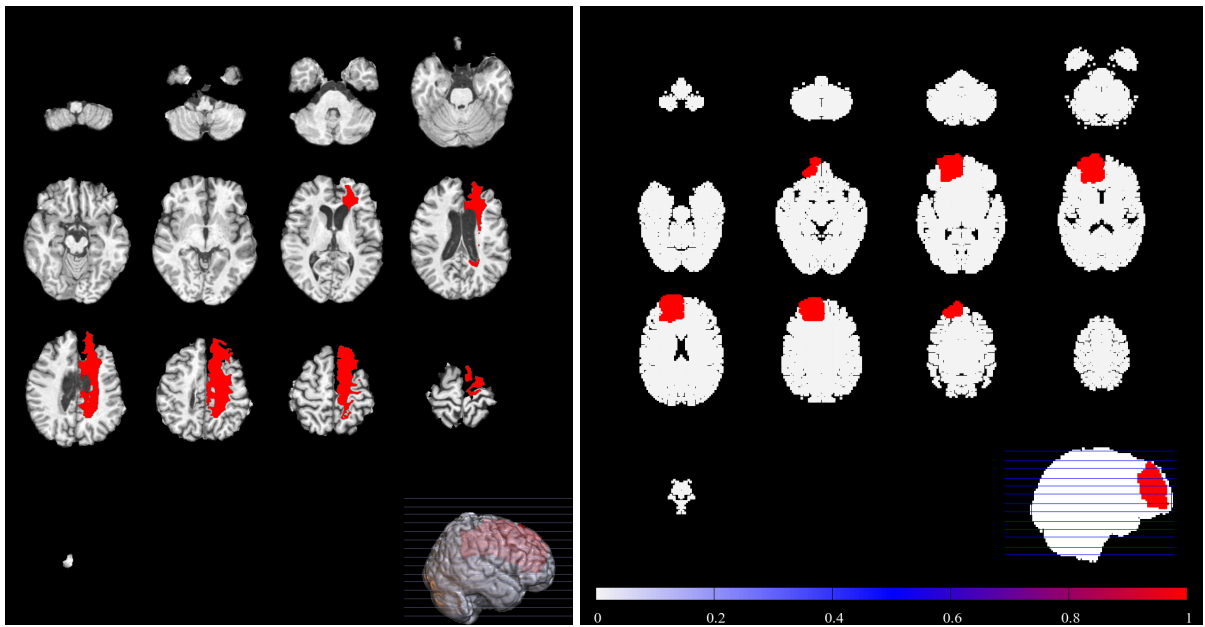

(c) PCA stroke:

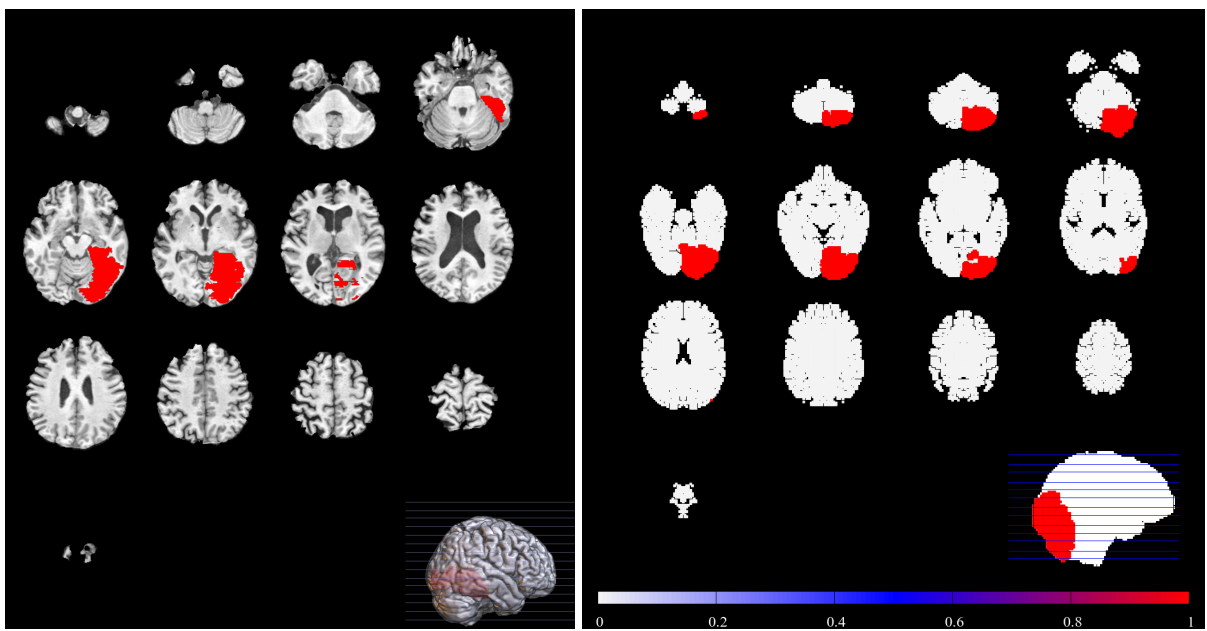

FIG 5: Maps showing the probability that emboli of particular sizes will arrive at a location in the brain. Small emboli are evenly distributed, whereas large emboli preferentially come to rest in the MCA territory of the *in silico* model. When receiving a distribution containing emboli of various sizes, emboli arrive preferentially in the posterior + superior MCA territory of the *in silico* model, but lesions are also common in the posterior and superior parts of the brain.

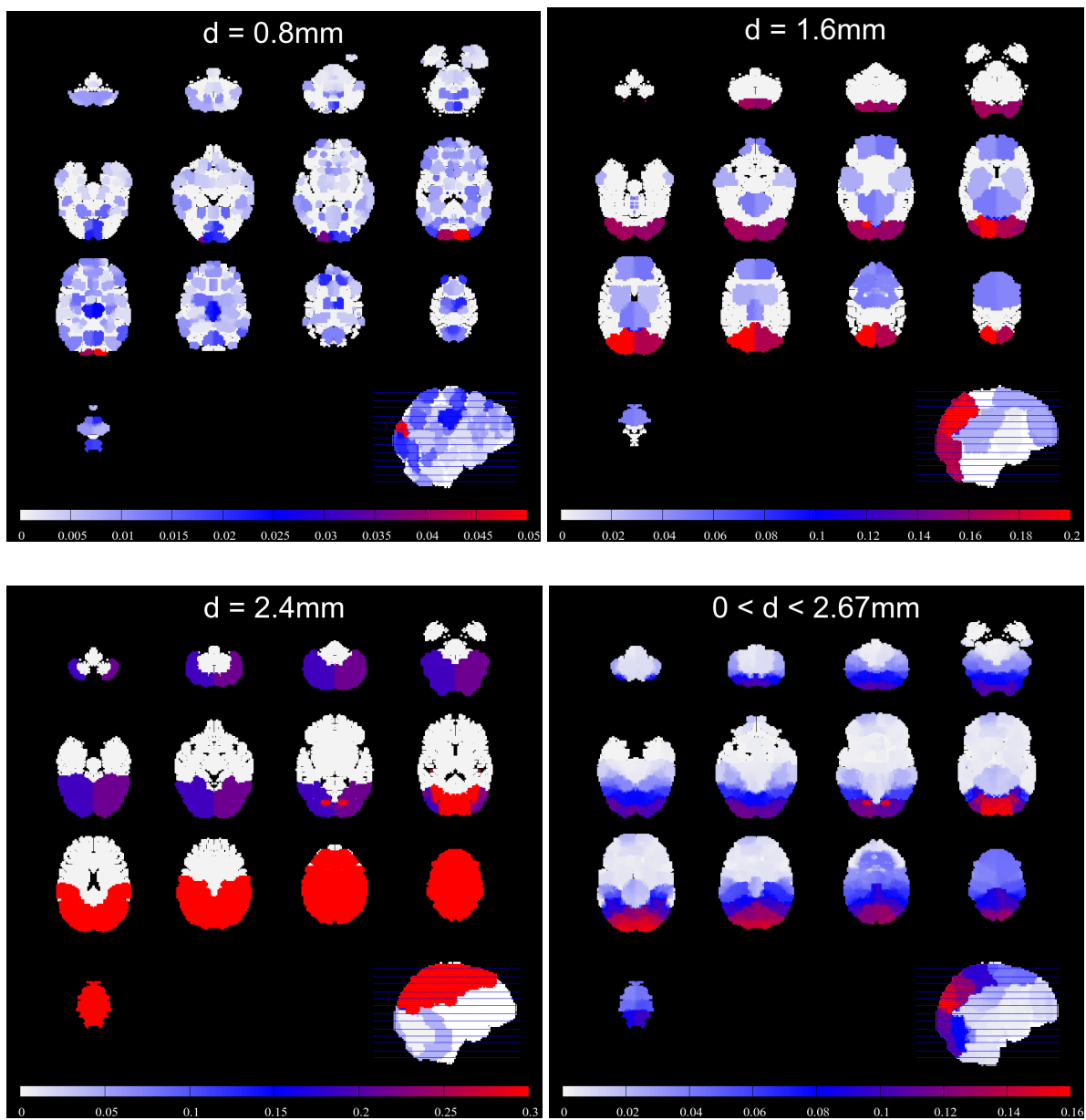